%% file: main.tex
\documentclass[10pt, sigconf, letterpaper, nonacm]{acmart}

\usepackage{algorithmic}
\usepackage{graphicx}
\usepackage{textcomp}
\usepackage{xcolor}
\usepackage{hyperref}
\usepackage{mathtools}

\usepackage{numprint}
\npthousandsep{,}

\usepackage{environ}
\usepackage{etoolbox}

\def\BibTeX{{\rm B\kern-.05em{\sc i\kern-.025em b}\kern-.08em
    T\kern-.1667em\lower.7ex\hbox{E}\kern-.125emX}}

\usepackage{enumitem}
\usepackage{xspace}
\usepackage{acro}
\acsetup{single}

\DeclarePairedDelimiter\abs{\lvert}{\rvert}%
\DeclarePairedDelimiter\norm{\lVert}{\rVert}%

\makeatletter
\let\oldabs\abs
\def\abs{\@ifstar{\oldabs}{\oldabs*}}
\let\oldnorm\norm
\def\norm{\@ifstar{\oldnorm}{\oldnorm*}}
\makeatother

\newtoggle{doPrintCommentsAll}
\NewEnviron{maybePrintAll}{%
  \iftoggle{doPrintCommentsAll}{\BODY}{}%
}

\toggletrue{doPrintCommentsAll}

\input{macros.tex}

\DeclareAcronym{DeFi}{
  short = DeFi,
  long  = Decentralized Finance,
}
\newcommand{\DeFi}{\ac{DeFi}\xspace}

\DeclareAcronym{CeFi}{
  short = CeFi,
  long  = Centralized Finance,
}

\DeclareAcronym{DApps}{
  short = dApps,
  long  = Decentralized Applications,
}

\DeclareAcronym{MEV}{
  short = MEV,
  long  = Miner/ Maximum Extractable Value,
}
\newcommand{\MEV}{\ac{MEV}\xspace}

\DeclareAcronym{BEV}{
  short = BEV,
  long  = Blockchain Extractable Value,
}

\DeclareAcronym{FaaS}{
  short = FaaS,
  long  = Front-running as a Service,
}

\DeclareAcronym{DApp}{
  short = DApp,
  long  = Decentralized Application,
}

\DeclareAcronym{IPS}{
  short = IPS,
  long  = Intrusion Prevention System,
}

\DeclareAcronym{EVM}{
  short = EVM,
  long  = Ethereum Virtual Machine,
}

\DeclareAcronym{BSC}{
  short = BSC,
  long  = BNB Smart Chain,
}

\DeclareAcronym{ABI}{
  short = ABI,
  long  = Application Binary Interface,
}

\DeclareAcronym{AMM}{
  short = AMM,
  long  = automated market maker,
}

\DeclareAcronym{OFAC}{
  short = OFAC,
  long  = Office of Foreign Assets Control,
}
\newcommand{\OFAC}{\ac{OFAC}\xspace}

\DeclareAcronym{SDN}{
  short = SDN,
  long  = Specially Designated Nationals and Blocked Persons,
}
\newcommand{\SDN}{\ac{SDN}\xspace}

\DeclareAcronym{DDoS}{
  short = DDoS,
  long  = Distributed Denial of Service,
}

\DeclareAcronym{DoS}{
  short = DoS,
  long  = Denial of Service,
}
\newcommand{\DoS}{\ac{DoS}\xspace}

\DeclareAcronym{PoW}{
  short = PoW,
  long  = Proof-of-Work,
}
\newcommand{\PoW}{\ac{PoW}\xspace}

\DeclareAcronym{PoS}{
  short = PoS,
  long  = Proof-of-Stake,
}
\newcommand{\PoS}{\ac{PoS}\xspace}

\DeclareAcronym{ZK}{
  short = ZK,
  long  = Zero Knowledge,
}

\DeclareAcronym{ZKP}{
  short = ZKP,
  long  = Zero Knowledge Proof,
}

\DeclareAcronym{SNARK}{
  short = SNARK,
  long  = Succinct Non-Interactive Argument of Knowledge,
}

\DeclareAcronym{RPC}{
  short = RPC,
  long  = Remote Procedure Call,
}
\newcommand{\RPC}{\ac{RPC}\xspace}

\DeclareAcronym{DEX}{
  short = DEX,
  long  = Decentralized Exchange,
}

\DeclareAcronym{FIFO}{
  short = FIFO,
  long  = First In First Out,
}

\DeclareAcronym{ICO}{
  short = ICO,
  long  = Initial Coin Offering,
}
\newcommand{\ICO}{\ac{ICO}\xspace}

\DeclareAcronym{TC}{
  short = TC,
  long  = Tornado Cash,
}
\newcommand{\TC}{\ac{TC}\xspace}

\newtheorem{theorem}{Theorem}

\title{Blockchain Censorship}
\author{Anton Wahrstätter\textsuperscript{1}, Jens Ernstberger\textsuperscript{2,8}, Aviv Yaish\textsuperscript{3}, Liyi Zhou\textsuperscript{4,8}, Kaihua Qin\textsuperscript{4,8},\\ Taro Tsuchiya\textsuperscript{5}, Sebastian Steinhorst\textsuperscript{2}, Davor Svetinovic\textsuperscript{1,6}, Nicolas Christin\textsuperscript{5},\\ Mikolaj Barczentewicz\textsuperscript{8}, Arthur Gervais\textsuperscript{8,9}}

\affiliation{
	\textsuperscript{1}Vienna University of Economics and Business
    \country{}
    \textsuperscript{2}Technical University of Munich
    \country{}
    \textsuperscript{3}The Hebrew University
    \country{}
    \\
	\textsuperscript{4}Imperial College London
    \country{}
    \textsuperscript{5}Carnegie Mellon University
    \country{}
    \textsuperscript{6}Khalifa University
    \country{}
    \textsuperscript{7}University of Surrey
    \country{}
    \\
	\textsuperscript{8}Berkeley Center for Responsible Decentralized Intelligence (RDI)
    \country{}
	\textsuperscript{9}University College London
}
\renewcommand{\shortauthors}{Wahrstätter et al.}

\thanks{This work was partially supported by the Ministry of Science \& Technology, Israel, the Federal Ministry of Education and Research of Germany (in the programme of “Souverän. Digital. Vernetzt.“. Joint project 6G-life, project identification number: 16KISK002), by the Carnegie Mellon CyLab Secure Blockchain Initiative, by a Nakajima Foundation Scholarship, and by an Ethereum Foundation grant.}

\begin{document}
\begin{abstract}
Permissionless blockchains promise to be resilient against censorship by a single entity. This suggests that deterministic rules, and not third-party actors, are responsible for deciding if a transaction is appended to the blockchain or not. In 2022, the U.S. \OFAC sanctioned a Bitcoin mixer and an Ethereum application, putting the neutrality of permissionless blockchains to the test.

In this paper, we formalize, quantify, and analyze the security impact of blockchain censorship. We start by defining censorship, followed by a quantitative assessment of current censorship practices. We find that $46\%$ of Ethereum blocks were made by censoring actors that intend to comply with OFAC sanctions, indicating the significant impact of OFAC sanctions on the neutrality of public blockchains.

We further uncover that censorship not only impacts neutrality, but also security. We show how after Ethereum's move to \PoS and adoption of \PBS, the inclusion of censored transactions was delayed by an average of $85\%$. Inclusion delays compromise a transaction's security by, e.g., strengthening a sandwich adversary. Finally, we prove a fundamental limitation of PoS and \PoW protocols against censorship resilience.
\end{abstract}

\maketitle

\section{Introduction}\label{sec:introduction}
Permissionless blockchains, such as Bitcoin and Ethereum, enable participants to interact with each other without the need for a trusted third party. In theory, these blockchains can be used by pseudonymous users without anyone being capable of censoring or seizing control of the network.

The use of pseudonymous addresses allows users to obfuscate their real identities, a popular feature among malicious actors~\cite{Weber2019,crawford2020,Christin_2013}.
Related works already study the use of blockchain applications for money laundering and other illicit purposes~\cite{Ruffing_2014, Stockinger_2021}. These malpractices attracted the attention of governments. Notably, the U.S.\ \OFAC included blockchain addresses in its \SDN list. Those subject to the U.S.\ jurisdiction are prohibited to interact with persons and property on the \SDN list. \OFAC sanctioned the cryptocurrency service provider Blender.io on May 5, 2022, for using its privacy-enhancing technology in a way that facilitated criminal money laundering. This was followed by the sanctioning of \TC on August 8, 2022, for the same reason. Blender.io is a centralized service for hiding Bitcoin (BTC) money flows, requiring users to trust those managing the service. In contrast, \TC is an autonomous and immutable smart contract application on Ethereum~\cite{pertsev2019tornado}. \TC has not been relatively significant to Ethereum in terms of the number of transactions or their value. Nevertheless, \OFAC sanctions against smart contract addresses are unprecedented and resulted in a significant reaction in the ecosystem, with reputable cryptocurrency providers restricting certain users from using their services~\cite{wright2022aave}.\\

\textbf{This paper. }
We begin by providing a holistic overview of blockchain censorship (\S~\ref{sec:overview}).
We focus on censorship on the consensus layer, as validator nodes are responsible for including transactions in a block, and censorship on the application layer, as smart contracts can prevent the successful execution of transactions included in a block (\S~\ref{sec:def}).

We quantitatively determine the extent of censorship due to \OFAC sanctions on Ethereum before (\S~\ref{subsection:prePBNS}) and after (\S~\ref{sec:postPbsCensorship}) its transition to \PoS (``the merge'').
We show that interactions with \TC's smart contracts declined by $84.3\%$ within two months following the sanctions. Furthermore, we provide evidence that Ethermine, which controlled $22\%$ of the hash rate in \PoW Ethereum, refused to include \TC interactions in their blocks. This resulted in a decrease of $200$ blocks ($\sim{33.4}\%$) containing \TC transactions per day. For post-merge Ethereum, we find that over a time frame of two months, at least $46\%$ of the total blocks were made by actors engaged in transaction censorship due to \OFAC sanctions.
At the application layer, we observed a spike in blocked users by $84.99\%$ in August 2022, the month of the introduction of the \OFAC sanctions.
On Bitcoin, we find that \OFAC sanctions prevented the Bitcoin mixer Blender.io from continuing to provide its centralized services (\S~\ref{sec:BitcoinBlender}).

We also study the implications of censorship on blockchain security (\S~\ref{sec:securityImplications}). We find that censorship delays the inclusion of \textit{both} censored and non-censored transactions by increasing their time in the memory pool (i.e., mempool) (\S~\ref{sec:confirmationLatency}).
Finally, we prove an impossibility result, which states that if $> 50\%$ of validator nodes \directly censor transactions, a \PoS blockchain cannot achieve censorship resilience (\S~\ref{sec:impossibility}).
To the best of our knowledge, we are the first to provide an empirical overview of applied censorship measures (\S~\ref{sec:quantification}) and associated security implications~(\S~\ref{sec:securityImplications}).
Thereby, our contributions are threefold:

\begin{itemize}
    \item We provide a definition of censorship for blockchains, capturing various system layers, and temporal features. We dissect the quantitative extent of censorship performed by Ethereum smart contracts, block builders, block proposers, and block relayers.
    
    \item We provide quantitative evidence of the historical transaction confirmation latency on Ethereum. We find that Ethereum's move to \PoS and \PBS has delayed the inclusion of both censored and non-censored transactions. E.g., the average inclusion delay for \TC transactions increased from~$15.8 \pm 22.8$ seconds in Aug.\ $2022$ to~$29.3 \pm 23.9$ seconds in Nov.\ $2022$. Increased confirmation latencies exacerbate sandwich attack risks.

    \item We prove that no \PoS (\PoW) protocol can achieve \textit{censorship-resilience}, if the censoring validators (miners) make up more than 50\% of the validator committee (hashing power).
\end{itemize}

\section{Background}\label{sec:background}
In the following, we provide the required background information on permissionless blockchains. 


\subsection{Permissionless Blockchains}
Permissionless blockchains build upon the premise of relying on a deterministic set of rules, instead of trusted parties, to determine the validity of a transaction. Bitcoin~\cite{nakamoto2008bitcoin} represents the first example of a permissionless blockchain that enables any entity to engage by creating transactions and broadcasting them to miners that eventually include them in a block that is appended to the blockchain. For the most part, Bitcoin transactions represent monetary flows between peers, though it is also possible to write arbitrary data onto the Bitcoin blockchain. Ethereum~\cite{wood2014ethereum} goes further than Bitcoin by allowing the deployment of arbitrary code, commonly referred to as \emph{smart contracts}, to the blockchain, which is then executed in a decentralized manner. Smart contracts gave birth to a thriving ecosystem of financial applications, \DeFi. Competitive trading on \DeFi emerged along with novel attacks~\cite{zhou2022sok}, such as sandwich attacks~\cite{zhou2021high}, more generally, front- and back-running~\cite{qin2022quantifying, qin2021empirical} exploiting transaction ordering for a financial gain. While Bitcoin relies on \PoW~\cite{bano2019sok}, Ethereum transitioned in September 2022 from \PoW to a \PoS leader election protocol in an event called ``the merge''~\cite{ethereum2022merge}. 

\begin{description}[leftmargin=0cm,labelindent=\parindent, itemsep=0.5em]
\item[Proposer/Builder Separation. ]
Shortly after the transition to \PoS, \PBS was introduced to Ethereum. \PBS separates the functions of creating new blocks and appending blocks to the blockchain. This is a direct response to problems associated with \MEV~\cite{daian2020flash, qin2022quantifying}, and supposedly should also enhance Ethereum's censorship-resistance~\cite{ethereum2023proposer}. \MEV extraction can negatively affect user experience~\cite{wang2022impact}, and more importantly, the underlying incentive structure of the blockchain, thereby harming blockchain security~\cite{zhou2021just, yaish2022blockchain,qin2022quantifying}.
In \PBS, the role of a ``validator'' (``miner'' in \PoW) is divided between  separate entities, namely ``block builders'' and ``block proposers'' (i.e. the validators themselves). In addition, ``relays'' were introduced, with the role of intermediating and establishing the required trust between block builders and proposers. Currently, it is optional for Ethereum validators to participate in \PBS and they can do so by using software called \href{https://github.com/flashbots/mev-boost}{MEV-Boost}. Validators are still free not to use MEV-Boost and to build blocks on their own.


\item[Privacy-Enhancing Technologies. ]
In blockchains such as Bitcoin and Ethereum, everybody can follow asset flows through the protocol~\cite{androulaki2013evaluating,victor2020address}. To compensate for this intrusion into financial privacy, privacy-enhancing technologies and services emerged. Commonly, although not always accurately, referred to as ``mixing services''~\cite{glaeser2022foundations}, they enable participants to obfuscate asset flows by either creating shared transactions together with other users or by routing assets through shared addresses. For Bitcoin, Blender.io is an example of an application that attempts to enhance privacy by allowing users to deposit their assets into a shared account together with other users and later withdraw them to a newly created, pseudonymous account. Unlike Blender.io, CoinJoin~\cite{maurer2017anonymous} wallets such as Wasabi or Samurai Wallet do not require users to trust a service operator.

\hspace{1em} On Ethereum, \TC represents the most prominent example of a privacy-enhancing application~\cite{pertsev2019tornado}. \TC allows users to deposit assets into a shared account and later withdraw the assets anonymously to a newly generated address, thereby preventing observers from tracing asset flows~\cite{le2021amr}.
This is achieved by relying on Zero-Knowledge Succinct Non-Interactive Argument of Knowledge (zk-SNARKS)~\cite{bitansky2012extractable}.
\TC offers different ``pools'' in which users can deposit assets of a fixed denomination, such as a $0.1$, $1$, $10$, or $100$ ETH. Users who deposit funds into a given denomination's contract can later withdraw the same amount from the respective pool without revealing their deposit address. The fixed deposit denomination is intended to make it difficult for observers to link deposits to withdrawals.

\hspace{1em} Governments suspect that the three privacy-enhancing technologies discussed above have been used by malicious actors such as the North Korean Lazarus Group to launder money and evade economic sanctions~\cite{park2021lazarus}.

\end{description}

\subsection{Rationales for Censorship}\label{sec:sanctions}

A blockchain's design goal to provide censorship-resistance may be challenged in practice by pressures for censorship. Blockchain actors who are capable of obstructing transactions submitted by users, or even of obstructing the finalization of blocks, may have a variety of reasons to do so.

Some of these reasons are, in a sense, \emph{external} (exogenous) to such actors and to the blockchain ecosystem in which they participate. The key case of that is pressures exerted by governments and the law. Other pressures are meaningfully \emph{internal} (endogenous): a desire to act ethically or a desire to profit economically.

Endogenous and exogenous reasons are in practice intertwined. For example, assume that in some jurisdictions it is unclear whether the law requires node operators to obstruct blockchain transactions involving addresses linked to a specific criminal organisation. In this situation, a node operator may decide to obstruct for many reasons, including (1) lowering their risk of legal liability, (2) genuine desire not to facilitate criminal activity, (3) economic motives (e.g., appearing as a responsible business operator to investors). Only the first reason is clearly exogenous, and even then, whether any operator will act on that reason, will depend on their tolerance for risk (an endogenous factor), especially given that the law is unclear. 

\begin{description}[leftmargin=0cm,labelindent=\parindent, itemsep=0.5em]
\item[Legal and Political Rationales. ]
Blockchain transactions are sometimes used for purposes that are criminalized in most jurisdictions, like hacking, theft, or payments facilitating crimes (e.g., for Child Sexual Abuse Materials~\cite{ndiaye2021cryptocurrency,cremona2019cybersecurity}, drugs and dark web markets~\cite{christin2022measuring}). Also, blockchain transactions are used for purposes prohibited for national security or humanitarian reasons, where there is less convergence across various jurisdictions. The key case of the latter is violations of economic sanctions laws. Targeting senders or addressees of such transactions may be challenging for law enforcement. Hence, from the perspective of prevention and deterrence of legally undesirable behavior, it may seem attractive to impose legal obligations to censor transactions. 

\item[The U.S.\ Economic Sanctions. ]
Under U.S. sanctions law, it is prohibited to engage in transactions with sanctioned entities, their property, and their interests in property~\cite{ofac2021guidance}. It is also prohibited to make ``any contribution or provision of funds, goods, or services by, to, or for the benefit of'' to sanctioned entities.~\cite{potus2015eo13694} 
The U.S.\ \OFAC maintains the Specially Designated Nationals and Blocked Persons (SDN) list, which has included blockchain addresses of sanctioned persons and organisations since $2018$.~\cite{ofac2021guidance} We refer to addresses included in the \SDN list as ``sanctioned addresses.'' On a technical note, Ethereum addresses (accounts) tend to prove more persistent than Bitcoin addresses (UTXO).~\cite{victor2020address}

\hspace{1em} We focus on two sanctions designations made by \OFAC in $2022$: Blender.io in May and Tornado Cash in August (re-designated in November)~\cite{treasury2022blenderio,treasury2022tornadoaugust,treasury2022tornadonovember}. In both cases, blockchain addresses were added to the \SDN list. Notably, in the \TC case, some \SDN-listed addresses refer to smart contracts without administrative functionality. This was the first time that smart contract addresses were added to the \SDN list. \OFAC later clarified that the open-source code of \TC smart contracts is not in itself sanctioned, only its instances deployed by the Tornado Cash organisation.~\cite{treasury2022faq1076}

\hspace{1em} The \TC sanctions motivated blockchain node operators to censor transactions involving addresses on the \SDN list (cf.\ Section~\ref{sec:quantification}). However, it is subject to debate whether censorship by blockchain validators, block builders, or relays, is required by law~\cite{seira2022base}. Moreover, \emph{if} those network participants are legally required to censor, it may be insufficient only to censor addresses on the \SDN list without attempting to censor unlisted addresses used by sanctioned entities, as \OFAC clarified that the \SDN list is not exhaustive in this respect~\cite{ofac2021guidance}. When we use the term ``OFAC-compliance'', we do so informally, referring to the likely rationale of the actor in question, while allowing for the possibilities that their actions are not legally required or that they are insufficient for compliance with U.S.\ sanctions law.

\end{description}

    
\section{Related Work and Censorship Examples}\label{sec:relatedworks}




\begin{description}[leftmargin=0cm,labelindent=\parindent, itemsep=0.5em]
\item[Defining Censorship. ]
Related contexts in which ``censorship'' has been defined, at least indirectly (negatively), include works on censorship-resistance (circumvention) in information systems in general~\cite{danezis2004economics,leberknight2010taxonomy,khattak2016sok,tschantz2016sok}, and works on censorship-resistance of blockchains in particular~\cite{thin2018formal,zhang2019lay,xian2019improved,silva2020impact,crain2021red,karakostas2022sok}. In previous works ``censorship'' has been understood as:
\emph{(a)} not including blocks with transactions to or from targeted entities~\cite{silva2020impact,crain2021red};
\emph{(b)} publicly announcing an intent to exclude future transactions of targeted entities, e.g., by \emph{feather forking}~\cite{zhang2019lay};
\emph{(c)} refusing to attest to a chain that contains transactions from or to a targeted entity~\cite{silva2020impact,thin2018formal}.

\hspace{1em} The first kind of censorship may apply to block contents or the entity that mined or proposed the block~\cite{thin2018formal}. Censorship of the second kind may also apply on the block content or the identity of the respective user.
The third point is specific to \PoS-based blockchains~\cite{neuder2021low}. 
We focus on selective censorship within a network (an information system), instead of censorship of entire networks~\cite{perng2005censorship}.

\item[Censorship Attacks. ]
The literature explores multiple attack vectors relevant to censorship, ranging from \DoS~\cite{biryukov2015bitcoin}, eclipse~\cite{heilman2015eclipse, gervais2015tampering}, routing~\cite{apostolaki2017hijacking}, to prefix hijacking~\cite{tran2020stealthier}. Focusing on censorship on the consensus layer, Miller~\cite{miller2013feather} introduced the \emph{feather forking} attack, where attackers with a minority of the hash-rate in a \PoW blockchain can censor transactions, which was later expanded upon by McCorry~\emph{et al.}~\cite{mccorry2018smart}, who propose methods to censor confirmed and unconfirmed transactions. Regarding the possibility of censorship at the network layer, Loe~\emph{et al.}~\cite{loe2019you} show that two methods to join a cryptocurrency network, DNS seeding and IP hard-coding, are vulnerable to censorship.

\item[Censorship Examples. ]
As part of an attack or due to legal obligations, certain participants may be ignored or even blocked by others. 
\RPC endpoints can prevent users from broadcasting their transactions, e.g., in March $2022$ the Ethereum \RPC endpoint Infura censored OFAC-sanctioned entities~\cite{nelson2022crypto}.
In the front-end, wallet applications have been implicated in censoring transactions~\cite{nelson2022crypto}, and similarly, the web applications of \DeFi projects have refused to engage with users who received funds from \TC~\cite{wright2022aave}.
At the consensus layer, it was reported that a mining pool suppressed the inclusion of \ICO transactions~\cite{eskandari2019sok}.
We note that a temporal delay in the execution of a transaction may entail significant financial implications for the censored entity~\cite{zhou2021high}.

\item[Preventing Censorship. ]
Zhang~\emph{et al.}~\cite{zhang2019lay} propose a  multi-metric evaluation framework for quantifying the attack resistance of \PoW-based blockchains, including against feather-forking attacks. 
Kostiainen~\emph{et al.}~\cite{kostiainen2022censorship} develop a censorship resistant and confidential payment channel that can be deployed to EVM-compatible blockchains. 
Le and Gervais~\cite{le2021amr} construct a reward-enabled censorship resilient mixer. Lotem~\emph{et al.}~\cite{lotem2022sliding} present a mechanism for on-chain congestion detection which can partially defend against censorship attacks. 
Karakostas~\emph{et al.}~\cite{karakostas2022sok} outline a systematic approach to measure blockchain decentralization and argue that centralization can jeopardize censorship-resistance in permissionless blockchain protocols.

\end{description}

We build upon prior research with the goal to provide a quantitative overview of censorship on public and permissionless blockchains. To the best of our knowledge, we are the first to quantify censorship imposed by different actors in the ecosystem and elaborate on its security implications.

\section{Overview of Censorship}\label{sec:overview}
We proceed to outline our system model and provide a definition of censorship on permissionless blockchains. 

\subsection{System Model} \label{sec:systemmodel}
We extend the system model of Zhou~\emph{et.\ al}~\cite{zhou2022sok}:

\begin{description}[leftmargin=0cm,labelindent=\parindent, itemsep=0.5em]
    \item[\textbf{Network Layer.}]
    In a \blockchain, validators form a \PTwoP network by following a set of rules which determine the communication interface, peer discovery as well as procedures for joining and exiting the network. 
    Messages are transmitted between network participants via e.g., gossip or dedicated communication channels.
    A user may include its message (or ``transaction'') in the blockchain by joining the \PTwoP network through a self-operated node or by relying on intermediary services (i.e., \RPC providers). 

    \item[\textbf{\ConsensusLayer.}]
    On the \consensusLayer, a fault-tolerant consensus algorithm ensures that validators in the \PTwoP network are in agreement on a shared state.
    In a blockchain, a newly proposed block is appended by the validator which is elected through a leader election protocol (e.g., \PoW).
    A block consists of transactions, where the node appending the block to the blockchain decides on the order of included transactions.
    Nodes are incentivized through a \textit{block reward}, paid for validating a block, and a \textit{transaction fee}, which is paid by the client.
    Each included transaction advances the shared network state, which is replicated by each validator. 
    
    \item[\textbf{Application Layer.}]
    \textit{Decentralized applications} (i.e., smart contracts), are smart contracts that maintain a state.
    A smart contract is defined by a set of functions that cause state transitions and can be invoked through a transaction.
    A smart contract can interact with other contracts through \emph{internal calls}. While there is no limit on the number of contracts a contract can interact with, blockchains specify an upper limit on the number of instructions a transaction can execute (e.g., the \emph{gas limit} in Ethereum).
    
    \item[\textbf{Auxiliary Services.}]
    Auxiliary services, are e.g., browser-based cryptocurrency wallets, user-interfaces of decentralized applications, and off-chain oracles. 
\end{description}


\subsection{Notation \& Terminology}
\label{section:notation}
In this work, we assume a single blockchain $\ledgerShort$ consisting of blocks $\blockShort$, where $\blockIdentifierShort$ corresponds to a \textit{block identifier}, with $\blockHeightShort$ corresponding to the block height.
We say that $\blockShort \in \ledgerShort$, if a block $\blockShort$ is included in the blockchain $\ledgerShort$.
The blockchain $\ledgerShort$ is maintained by a set of $\numberValidators$ validators, which agree upon the current state of $\ledgerShort$ through a \stateMachineReplication protocol $\stateMachineReplicationShort$. The protocol $\stateMachineReplicationShort$ receives as an input a set of transactions $\transactionsShort$, and outputs the ordered ledger of transactions $\ledgerShort$. 
Let $\securityParameter$ be a security parameter that determines the finality of $\ledgerShort$. Then, we denote $\finalityTime$, a polynomial function in $\securityParameter$, as the \textit{finality delay}.
Based on this notation, we define \textit{transaction inclusion} as follows:
\begin{definition}[Transaction Inclusion]\label{def:inclusion}
    A transaction $\transactionShort$ received by a validator at time $t$ is \textit{included} in $\ledgerShort$ by the \stateMachineReplication protocol $\stateMachineReplicationShort$, if $\transactionShort \in \blockShort \mid \blockShort \in~\ledgerShort$ at time $t' > t + \finalityTime$.
\end{definition}

Further, we denote the address of an account maintained through $\ledgerShort$ as $\addressShort$. We intentionally do not differentiate between externally owned accounts and smart contracts, as this abstraction is irrelevant with regard to censorship.
When it comes to preventing censorship, we differentiate between censorship \textit{resistance} and censorship \textit{resilience}. 
Censorship resistance describes a technology that prevents protocol participants from censoring at all (e.g., confidential ``to'' addresses), whereas censorship resilience describes that the act of censorship is possible for an individual, but the respective system is resilient against it.

\subsection{Definition of Censorship}\label{sec:def}

Censorship is a broad term, which may apply in any of the system layers as introduced in Section~\ref{sec:systemmodel}.
To clarify the notion in the context of a blockchain, we set out to synthesize existing notations in formal definitions.
We begin by defining censorship on the consensus layer. 
\begin{description}[style=unboxed,leftmargin=0cm,labelindent=\parindent, itemsep=0.5em]
    \item[\textbf{Consensus Layer Censorship. }]
    Censorship on the consensus layer may either be enforced \textit{\directly} or \textit{\indirectly}.
    For example, a validator may enforce \direct censorship by refusing to broadcast a received transaction, sign an attestation, or include a transaction in a block (cf. \S~\ref{sec:quantification}).
    Alternatively, an external entity may \indirectly enforce censorship by preventing timely transmission of messages, or occupying validator nodes through targeted \DoS attacks. Hence, censorship on the consensus layer can also \indirectly originate from the network layer or application layer.
    Therefore, we focus our definition of censorship on the consensus layer on the \textit{intent} of a protocol participant to obstruct the inclusion of a transaction in a blockchain. 
    \begin{definition}[Strict Censorship]\label{def:censorship}
    A transaction is censored if a protocol participant intentionally obstructs the inclusion of a transaction,
    such that $\transactionShort \notin \blockShort \mid \blockShort \in \ledgerShort$.
    \end{definition}
     
     
     Furthermore, we identify a subtle variant of censorship, where transactions are included with a delay.
     
    \begin{definition}[Weak Censorship]\label{def:weakcensorship}
    A transaction is censored if an actor intentionally obstructs the inclusion of a transaction in the next possible block, such that a transaction $\transactionShort$, received at block height $h$, does not get included in a block $\blockShort$ at block height $h' = h+1$, thus $\transactionShort \in \blockShort \mid \blockShort \in \ledgerShort$, yet $h_{\blockShort} < h'_{\blockShort}$.
    
   \end{definition}

    \hspace{1em} Definition~\ref{def:censorship} and Definition~\ref{def:weakcensorship} follow related works in distinguishing \emph{censorship} from \emph{ordering} of transactions~\cite{buterin2021sok,qin2022quantifying,chitra2022improving,auer2022miners,heimbach2022sok}. 
    As transactions are only ordered once decided that they are included in a block, we do not treat it as ``censorship'' when block builders order (``re-order'') transactions differently than expected by the senders of these transactions (e.g., in Ethereum there may be an expectation of ordering only according to the fees paid by each transaction, which is the default behaviour~\cite{gafni2022greedy}). 
    We note, however, that intentional re-ordering may result in the transaction failing, or in a lower economic gain for the user (e.g., due to front-running~\cite{qin2022quantifying}).


    \item[\textbf{Application Layer Censorship. }]
    As previously defined, ``strict censorship'' cannot be enforced \directly by the application layer, i.e. by smart contracts, because they cannot \directly affect inclusion or finalization of blocks and transactions at the consensus layer.
    However, a smart contract can \indirectly influence blockchain censorship, by incentivizing inclusion or exclusion of transactions and blocks or incentivizing retroactive forking of a blockchain \cite{mccorry2018smart,winzer2019temporary,naumenko2022txwithhold}. 
    Besides \indirectly influencing the consensus layer, smart contracts can enforce \direct censorship by preventing the successful execution of transactions that are included in a block.
    We define \textit{smart contract censorship} as follows.
    
    \begin{definition}[Smart Contract Censorship]\label{def:contractcensorship}
        A transaction $\transactionShort$ is censored by a smart contract, if $\transactionShort \in \blockShort$, where $\blockShort \in \ledgerShort$ is blocked by the state $\stateShort_{i}$, s.t.\ further state transitions $\stateShort_{i} \xrightarrow{} \stateShort_{i+1}$ are blocked by the respective contract.
    \end{definition}

    An example of smart contract censorship is a block list, which prevents an account with address $\addressShort$ from successfully executing a transaction that interacts with the block listing smart contract (cf. Section~\ref{sec:quantification}). 
    
\end{description}

\section{Censorship Quantification}\label{sec:quantification}
In the following, we provide an empirical quantification of censorship on Ethereum and Bitcoin. 
We distinguish between pre- and post-merge Ethereum because the consensus mechanism impacts censorship.


\subsection{Data Collection}\label{sec:data}
We collect data about the OFAC-sanctioned applications \TC and Blender.io starting from the 1st of January 2021 00:00:00 UTC until the 15th of November 2022 23:59:59 UTC.

\begin{description}[style=unboxed,leftmargin=0cm,labelindent=\parindent, itemsep=0.5em]
\item[Blender.io Data. ]
For data on Blender.io, we set up a local Bitcoin node and parse the raw data files. We filter for transactions from and to the sanctioned addresses of Blender.io using the addresses listed in OFAC's SDN list. 

\item[\TC Data. ]
For collecting Ethereum application layer data, we connect to an \href{https://infura.io/}{\RPC provider Infura} and leverage the \href{https://etherscan.io/apis}{Etherscan API}. This includes event logs broadcast by sanctioned \TC contracts. The event logs indicate that a user has either deposited or withdrawn funds from a \TC contract. We include all existing \TC pool-contracts in all denominations (cf.\ Table~\ref{tab:tc-contracts}).
Notably, we did not include all sanctioned addresses, instead, we focus on deposits and withdrawal to the known \TC pool contracts. This means that, e.g., the \TC Gitcoin grant contract, contracts deployed on Layer-2 solutions such as Polygon or Arbitrum, or contracts only existing on an Ethereum testnet have been ignored. However, we capture most of the traffic from sanctioned entities because most users interact with the ETH-denominated contracts deployed to the Ethereum mainnet.
In total, our data set has \numprint{273,403} entries, each representing either a \TC deposit or withdrawal, that have been included in \numprint{236,868} distinct blocks.

\begin{table}[!t]
\footnotesize
\centering
\caption{\TC contracts we capture in this work.}
\label{tab:tc-contracts}
\begin{tabular}{c c}
\toprule
\TC Contract &                       Address \\
\midrule
  0.1 ETH  & 0x12D66f87A04A9E220743712cE6d9bB1B5616B8Fc \\
    1 ETH  & 0x47CE0C6eD5B0Ce3d3A51fdb1C52DC66a7c3c2936 \\
   10 ETH  & 0x910Cbd523D972eb0a6f4cAe4618aD62622b39DbF \\
  100 ETH  & 0xA160cdAB225685dA1d56aa342Ad8841c3b53f291 \\
 0.1k DAI  & 0xD4B88Df4D29F5CedD6857912842cff3b20C8Cfa3 \\
   1k DAI  & 0xFD8610d20aA15b7B2E3Be39B396a1bC3516c7144 \\
  10k DAI  & 0x07687e702b410Fa43f4cB4Af7FA097918ffD2730 \\
 100k DAI  & 0x23773E65ed146A459791799d01336DB287f25334 \\
  5k CDAI  & 0x22aaA7720ddd5388A3c0A3333430953C68f1849b \\
 50k CDAI  & 0x03893a7c7463AE47D46bc7f091665f1893656003 \\
500k CDAI  & 0x2717c5e28cf931547B621a5dddb772Ab6A35B701 \\
  5m CDAI  & 0xD21be7248e0197Ee08E0c20D4a96DEBdaC3D20Af \\
 100 USDC  & 0xd96f2B1c14Db8458374d9Aca76E26c3D18364307 \\
  1k USDC  & 0x4736dCf1b7A3d580672CcE6E7c65cd5cc9cFBa9D \\
 10k USDC  & 0xD691F27f38B395864Ea86CfC7253969B409c362d \\
 100 USDT  & 0x169AD27A470D064DEDE56a2D3ff727986b15D52B \\
1000 USDT  & 0x0836222F2B2B24A3F36f98668Ed8F0B38D1a872f \\
 10k USDT  & 0xF67721A2D8F736E75a49FdD7FAd2e31D8676542a \\
100k USDT  & 0x9AD122c22B14202B4490eDAf288FDb3C7cb3ff5E \\
 0.1 WBTC  & 0x178169B423a011fff22B9e3F3abeA13414dDD0F1 \\
   1 WBTC  & 0x610B717796ad172B316836AC95a2ffad065CeaB4 \\
  10 WBTC  & 0xbB93e510BbCD0B7beb5A853875f9eC60275CF498 \\
\bottomrule
\end{tabular}
\end{table}

\item[Ethereum Ecosystem Data. ]
We collect data on the different ecosystem participants, such as, miners, block proposers, block relayers and block builders.
For consensus layer data, we use the \href{https://beaconcha.in/}{beaconcha.in} API to collect information about \PoS slots and the validators and proposers involved in each. For information on external actors such as block builders and relay operators, we use the Relay Data API. The \href{https://flashbots.github.io/relay-specs}{ProposerPayloadsDelivered API endpoint} enables us to retrieve information on the parties involved in \PBS. In particular, we are interested in the blocks that block builders deliver to proposers. We connect to every existing relay provider by November 2022 including Flashbots, BloXroute, Blocknative, Manifold, Eden, and Relayooor. Summarizing, our final data set contains $443,831$ blocks, which includes every block since \PBS has been launched until the 15th of November 2022 23:59:59 UTC. We count a total of $256,078$ ($57.70\%$) blocks that were built by an external block builder. Consequently, a total of $187,754$ ($42.30\%$) blocks were built by the respective proposer locally. Finally, we use the \href{https://etherscan.io/labelcloud}{Label Word Cloud} to map addresses to identified entities. 


\item[\OFAC \SDN List. ]
In total, at the time of writing, OFAC's \SDN list includes $132$ Ethereum addresses. $90$ ($68\%$) of the sanctioned addresses belong to the privacy tool \TC. The remaining addresses are believed to belong to entities such as dark web criminals or hacking groups. Notably, in contrast to \TC, other SDN-listed Ethereum addresses refer to Externally Owned Accounts (EOAs) operated by end users, whereas most of the \SDN-listed \TC addresses are smart contracts, over which no Ethereum user has unrestricted control. Note that Ethereum testnet addresses are also sanctioned (e.g., Goerli testnet), while we focus on the Ethereum mainnet.

\end{description}


In the following sections, we start by identifying the effects of the sanctions on the \TC contracts by assessing their immediate impact on user engagement. Second, we focus on the effects of the sanctions on the individual validators. Third, we assess the impact of the sanctions on the distinct participants of the ecosystem. Thus, we distinguish between block builders, proposers, and relayers.

\subsection{Pre-\PBS Consensus-Layer Censorship}
\label{subsection:prePBNS}
\begin{figure}[!t]
\centering
\includegraphics[width=\columnwidth]{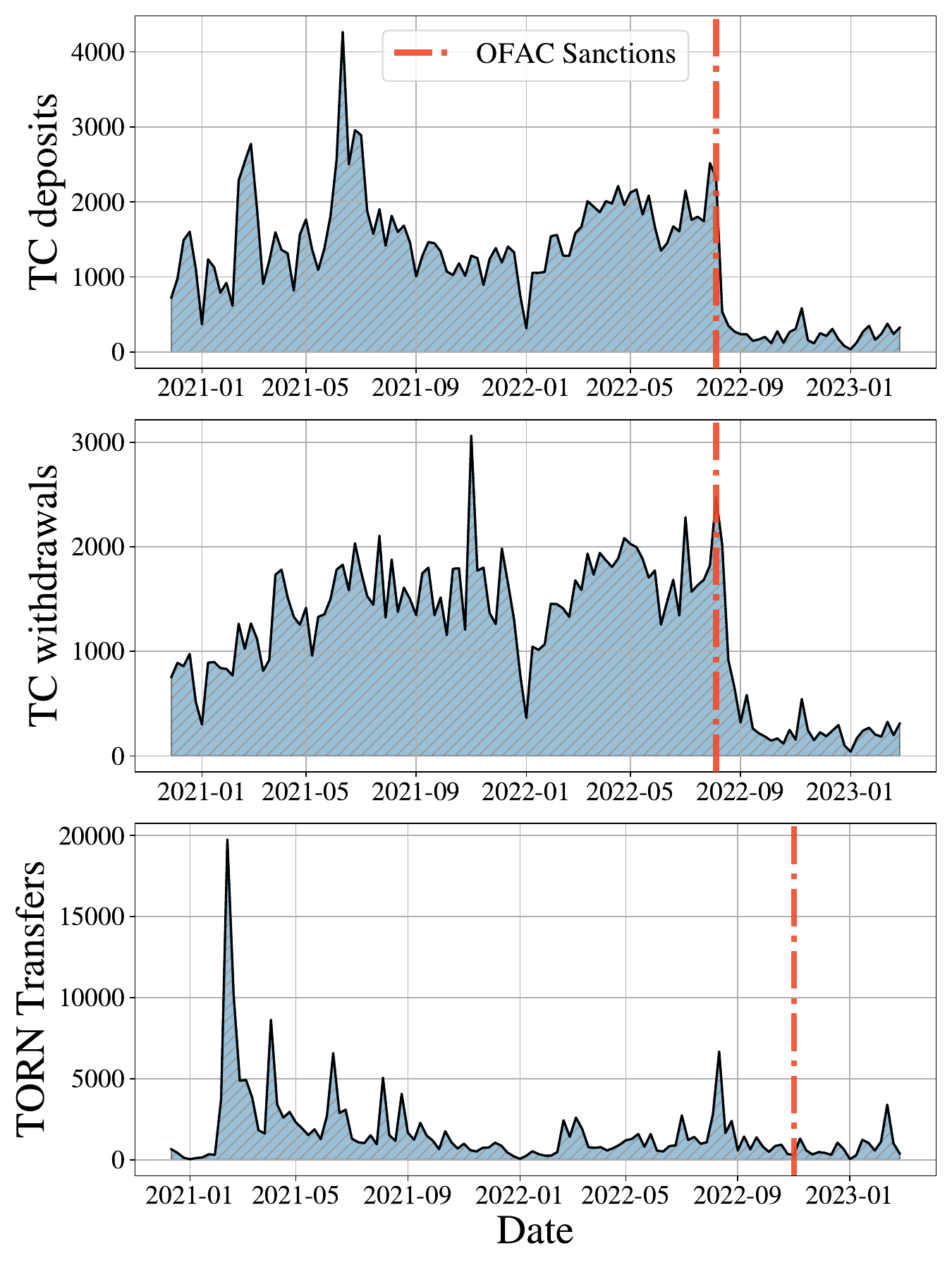}
\vspace{-2em}
\caption{\TC Interactions per week: \TC deposits and withdrawals over time, accounting for all deployed \TC contracts in every denomination from the 1st of October 2021 to November 15th, 2022.}
\label{fig:tornado-interactions}
\end{figure}

To illustrate the effects of \OFAC's sanctions, we show in Figure~\ref{fig:tornado-interactions} the number of interactions with \TC contracts over time, through the number of weekly deposits and withdrawals. While the weekly deposits and withdrawals reached over \numprint{2000} before the sanctions, \TC's activity afterward reduced by ten-fold to about \numprint{200} deposits and withdrawals per week. As of the enactment of the sanctions, we observe a decline in interactions with \TC contracts. For the month of October 2022, a total of \numprint{1630} interactions were observed, compared to \numprint{16347} interactions in July 2022. However, notably, the number of interactions has never dropped to zero.

Such a reduction in activity deteriorates \TC's anonymity set because users depend on other users for achieving sustainable privacy~\cite{wang2023how}. \TC relies on network effects, meaning that the more users participate, the greater the privacy gain for the individual. Smaller anonymity sets increase the risks of users being deanonymized through side-channels.

A reasonable explanation for the decrease in interactions with \TC's contracts is that due to the sanctions, the \TC website was promptly taken off-line~\cite{milton2022crypto}. Consequently, users could only interact with the respective contracts without using any interface, which may not have been feasible for a majority of the users. In addition, the open source Github repository that hosts the \TC code was temporarily taken offline, preventing users from redeploying the front-end.
\href{https://www.circle.com/en/}{Circle}, the company issuing the USDC stablecoin, froze all USDC tokens inside the \TC contracts. As a result, the owners of those assets can no longer move their funds.

\begin{figure}[!t]
\centering
\includegraphics[width=\columnwidth]{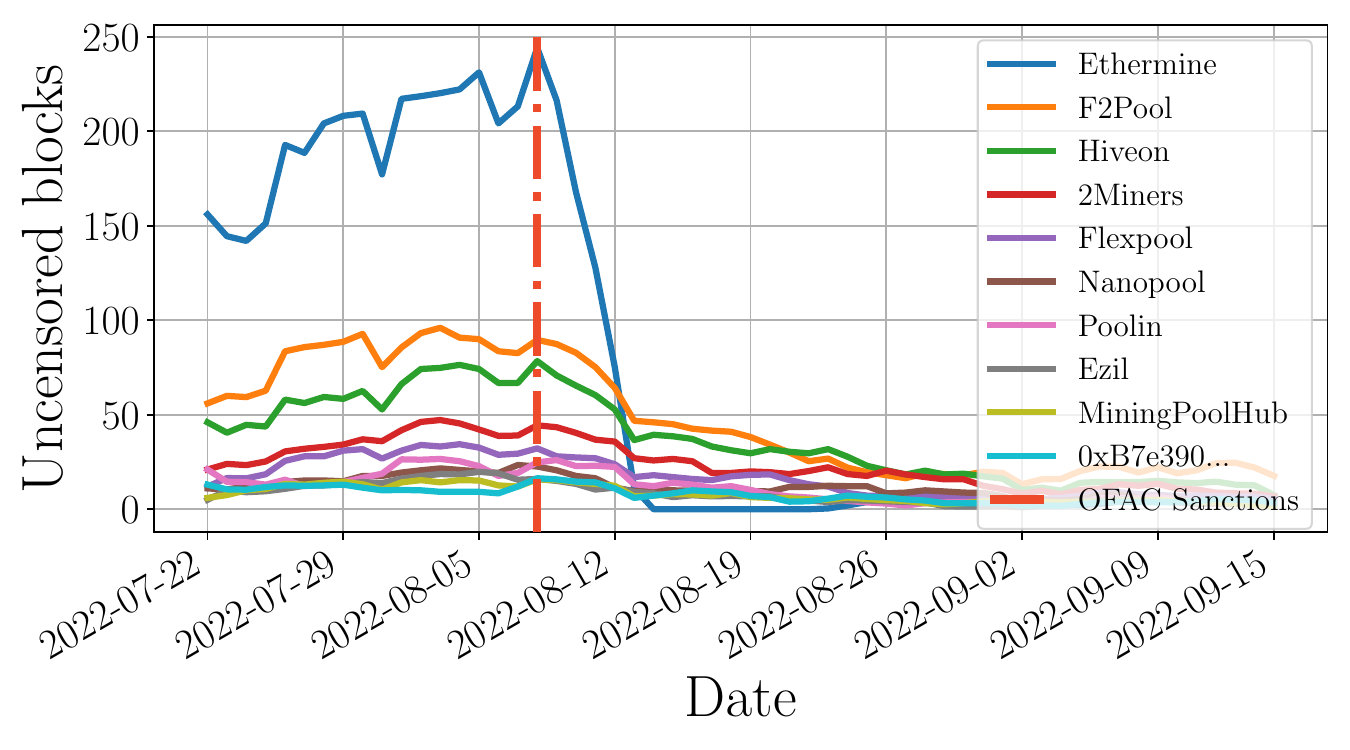}
\vspace{-2em}
\caption{Blocks containing \TC transactions by the top 10 miners of uncensored blocks from July 22, 2022 until the transition to \PoS on September 15, 2022 (5 day average).}
\label{fig:tornado-miners}
\end{figure}

\begin{description}[style=unboxed,leftmargin=0cm,labelindent=\parindent]
\item[How Miners React to Sanctions. ]
Shifting the focus to the largest miners, that eventually decide upon the inclusion of \TC transactions into their blocks, we visualize the number of uncensored blocks over time from July 1st, 2021 to September 15th, 2022 in Figure~\ref{fig:tornado-miners}. We observe a decrease in uncensored blocks for the $10$ largest miners, which is partly an expected consequence resulting from the overall decrease in TC transactions. Nevertheless, Figure~\ref{fig:tornado-miners} indicates that the decline has been more pronounced for Ethermine compared to other miners. Before the sanctions, we observe on average $608$ ($8.5\%$) blocks containing uncensored transactions per day. Prior to the sanctions, on average $203$ uncensored blocks per day were built by Ethermine, representing $\sim{33.4}\%$ of the total number of uncensored blocks per day. After the sanctions, the number of uncensored blocks built by Ethermine decreased to $\sim{21}$ blocks per day, which yields a reduction of almost $90\%$. For the remaining miners, we observe a decrease of uncensored blocks between $50\%$ and $65\%$, a significantly smaller decline, while no miner completely ceased including \TC transactions.
\end{description}

\subsection{Post-\PBS Consensus-Layer Censorship}\label{sec:postPbsCensorship}
On September 15th, 2022, $38$ days after the \TC sanctions, Ethereum transitioned to \PoS and partially adopted \PBS, adding new intermediaries to the ecosystem. Block builders, block proposers, and relayers all have their own responsibilities, as well as their own ways to censor the Ethereum blockchain. The adoption of \PBS has grown and until the $15^{th}$ of November 2022, $58\%$ of all blocks were built by third-party block builders.
We therefore divide the following section among the mentioned participants and analyze the degree of censorship for each of them separately. 

\begin{figure}[!t]
\centering
\includegraphics[width=\columnwidth,trim=4 4 4 4,clip]{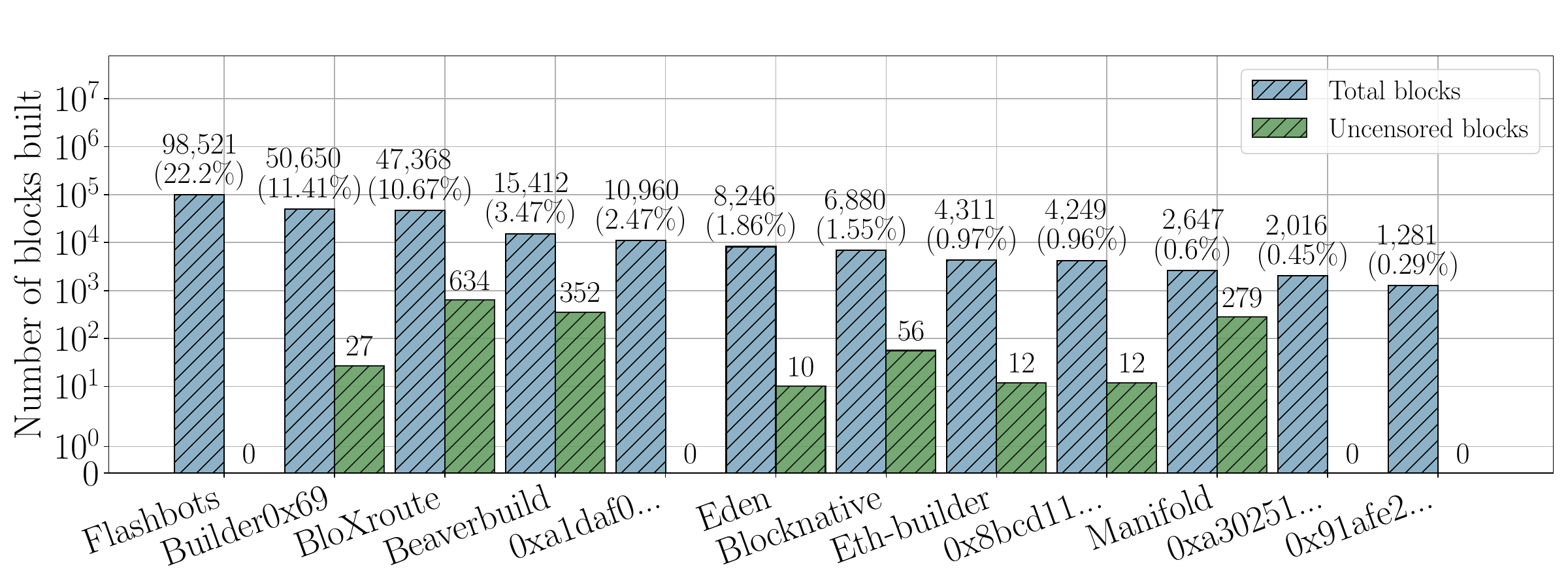}
\vspace{-2em}
\caption{Block builders on the Ethereum blockchain since the activation of \PBS at block height 15,537,940 (Sep-15-2022 08:33:47 AM) until block height 15,978,869 (Nov-15-2022 12:59:59 UTC). Uncensored blocks represent blocks containing interactions with \TC contracts.}
\label{fig:builder}
\end{figure}

\begin{figure}[!t]
\centering
\includegraphics[width=\columnwidth,trim=4 4 4 4,clip]{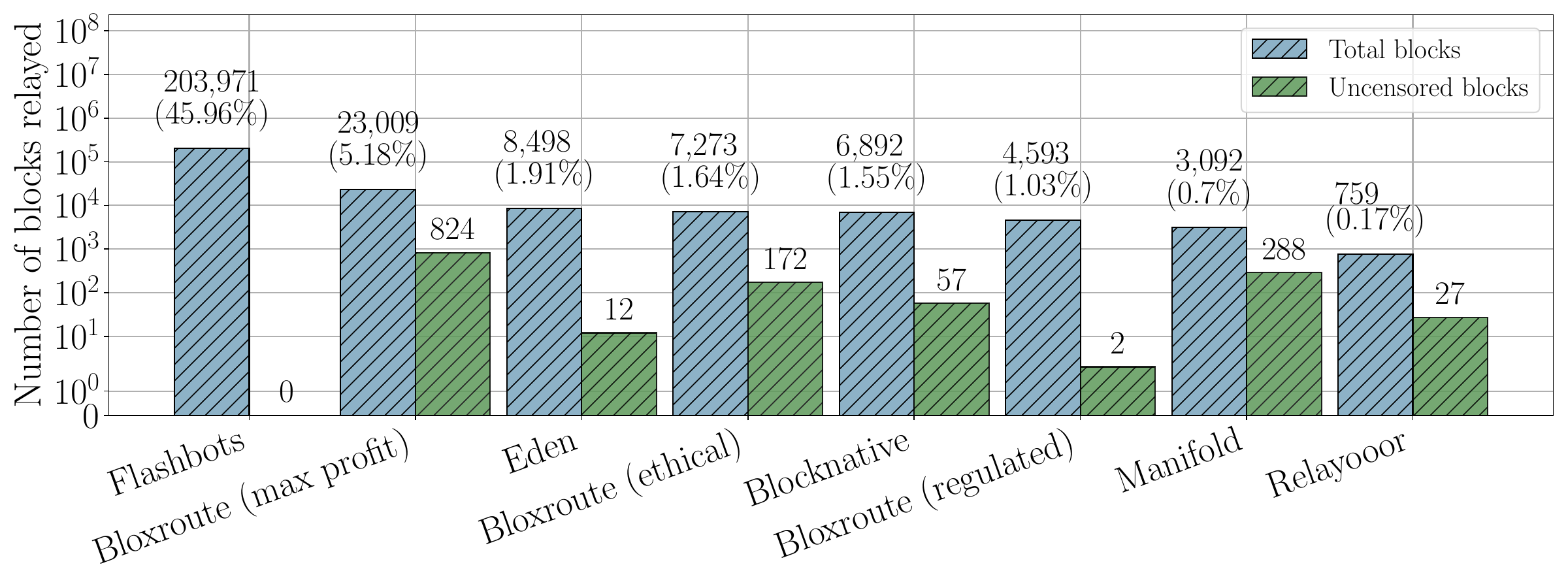}
\vspace{-2em}
\caption{Block relayers on the Ethereum blockchain since activation of \PBS at block height 15,537,940 (Sep-15-2022 08:33:47 AM) until block height 15,978,869 (Nov-15-2022 12:59:59 UTC). Uncensored blocks represent blocks containing interactions with the \TC contracts.}
\label{fig:relayers}
\end{figure}

\begin{description}[style=unboxed,leftmargin=0cm,labelindent=\parindent, itemsep=0.5em]

\item[Block Builder Censorship. ]
Starting with external block builders, who take bundles of transactions, construct blocks and pass them to block proposers, we display the ten largest block builders along with their total number of blocks proposed in Figure \ref{fig:builder}. We add the total number of uncensored blocks to reveal potential censorship practices.

\hspace{1em} As depicted in Figure \ref{fig:builder}, Flashbots' block builders are the most successful, as measured by the number of blocks they created. 
In total, Flashbots' builders are responsible for $\sim{22.2}\%$ of all blocks created between the \PoS transition and the $15^{th}$ of November 2022. This culminates to \numprint{97324} blocks in that timeframe. The builders of Builder0x69 are the second most successful block builders with a total of \numprint{50650} ($\sim{11.41}\%$) blocks, followed by the BloXroute builders that account for \numprint{47368} ($\sim{10.67}\%$) blocks and Beaverbuild with \numprint{15412} ($\sim{3.47}\%$) blocks.

\hspace{1em} Our results suggest that the four largest block builders of the Ethereum network engage in censoring by not including deposits \emph{to} and withdrawals \emph{from} the \TC contracts. The same applies to one of the BloXroute builders, accounting for $2.2\%$ of the total number of blocks built, as well as the anonymous builder with the public key 0xa1daf0..., who is responsible for $2.5\%$ of the total number of blocks.

\hspace{1em} Among the most successful builders displayed in Figure~\ref{fig:builder}, we find empirical evidence that only three of them incorporate \TC deposits and withdrawals into their blocks. Two of these non-censoring builders belong to BloXroute and the remaining one to Beaverbuild.

\item[Block Proposer Censorship. ]
Shifting the focus to block proposers, we visualize the most successful in Figure~\ref{fig:validators}. The staking pool Lido is the most successful group of block proposers between the launch of \PBS and the $15^{th}$ of November 2022, proposing \numprint{119068} ($\sim{26.83}\%$) valid blocks. The $2^{nd}$, $3^{rd}$ and $4^{th}$ most successful block proposers are the exchanges Coinbase, Kraken, and Binance with a total of \numprint{62964}, \numprint{36339} and \numprint{29317} blocks respectively. Additionally, we identify staking pools such as Staked.us, Figment, Rocketpool, and Stakefish as among the most successful block proposers.

\begin{figure}[!t]
\centering
\includegraphics[width=\columnwidth,trim=4 4 4 4,clip]{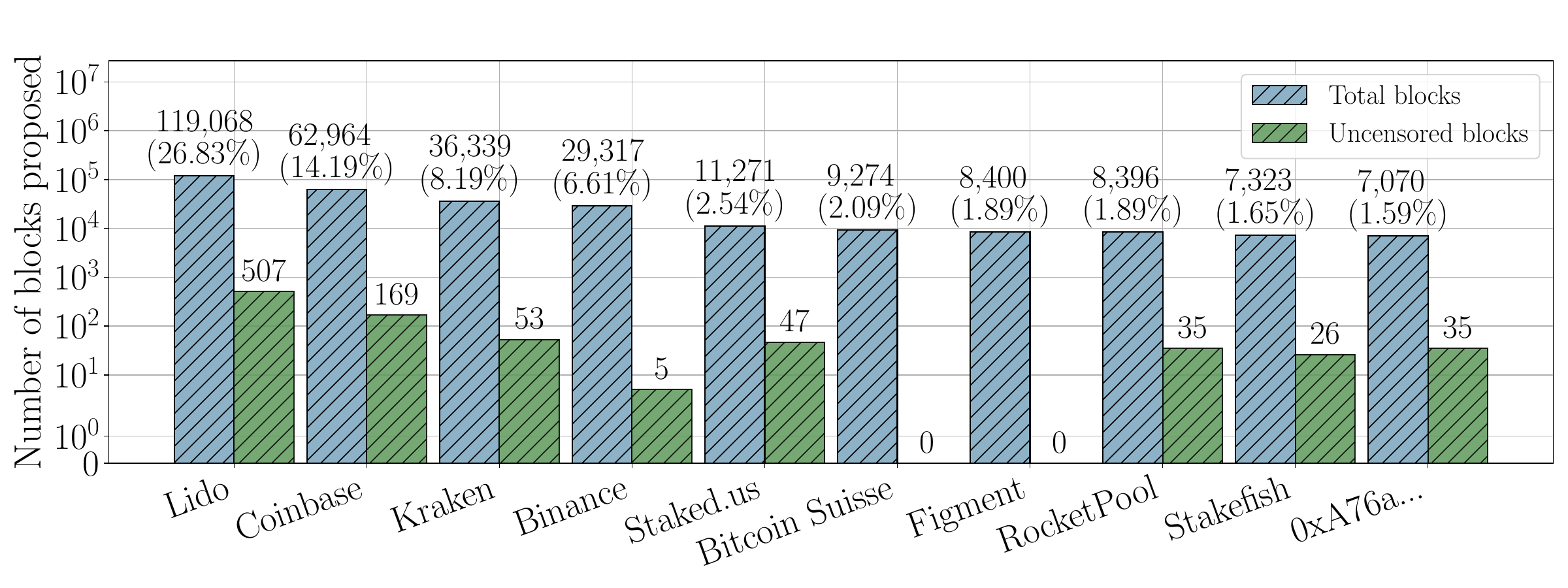}
\vspace{-2em}
\caption{Block proposers on the Ethereum blockchain between activation of \PBS at block height 15,537,940 (Sep-15-2022 08:33:47 AM) until block height 15,978,869 (Nov-15-2022 12:59:59 UTC). Uncensored blocks represent blocks containing transactions that interact with \TC contracts.}
\label{fig:validators}
\end{figure}

\hspace{1em} Focusing on the number of uncensored blocks, we find that among the ten block proposers displayed, Bitcoin Suisse and Figment never included deposits and withdrawals to \TC's contracts within the analyzed period. Together, both proposers account for almost $3\%$ of the total number of blocks proposed. 
Consequently, we can probabilistically infer that both entities engage in censoring by excluding \TC transactions from their blocks.
At this stage, it is important to emphasize that block proposers, who have chosen to adopt \PBS with MEV-Boost, rely to a large extent on blocks supplied to them by external block builders.

\item[Block Relayer Censorship. ]
Third, we analyze block relayers, who intermediate between block builders and block proposers.
In Figure \ref{fig:relayers}, we visualize the existing block relayers and the number of blocks forwarded to block proposers that were eventually added to the blockchain. Relayers may simulate blocks received from builders and thus censor the network by exclusively forwarding to block proposers blocks that do not include interactions with \SDN addresses.

\hspace{1em} At the time of writing, $85\%$ of blocks pass one of the depicted relayers. On Ethereum, there are $8$ relay services, three operated by BloXroute. Flashbots relays $79\%$ of blocks. BloXroute's ``max profit'' relay comes second with a market share of $8.9\%$. The remaining relayers have a market share between $3.3\%$ and $0.3\%$. Compared to the total number of blocks since the activation of \PBS, the Flashbots relayed $\sim{46}\%$ of the blocks proposed.
Similarly to what we observe with their block builders, Flashbots' relay has consistently not forwarded blocks including \TC transactions, while other relayers forward blocks with \TC transactions.
\end{description}

Concluding, we find that \PBS impacts censorship on Ethereum. Block builders and block relayers impose censorship on proposers who are using MEV-Boost. \PBS enables block proposers to boost their profits by additionally capturing the \MEV in the proposed blocks. As the most successful block builders and block relayers censor \TC transactions, block proposers must decide whether to adopt censorship or exclusively connect to a non-censoring relay. 

While the censoring block proposers Bitcoin Suisse and Figment both use MEV-Boost, for Bitcoin Suisse we find that only $0.28\%$ of their \numprint{9271} blocks were built by external \PBS block builders. 
Blocks proposed by Bitcoin Suisse were relayed by Blocknative, BloXroute (``max profit''), BloXroute (``regulated'') and Eden. Notably, while BloXroute (``max profit'') does not censor \TC transactions, there were no \TC transactions in the blocks that were relayed by BloXroute (``max profit'') and eventually proposed by Bitcoin Suisse. For Figment, $96.8\%$ of the \numprint{8400} blocks were built by third-party block builders from censoring relayers (i.e., Flashbots and BloXroute (``regulated'')).

\subsection{Application Layer Censorship}
\begin{figure}[!t]
\centering
\includegraphics[width=\columnwidth,trim=4 4 4 4,clip]{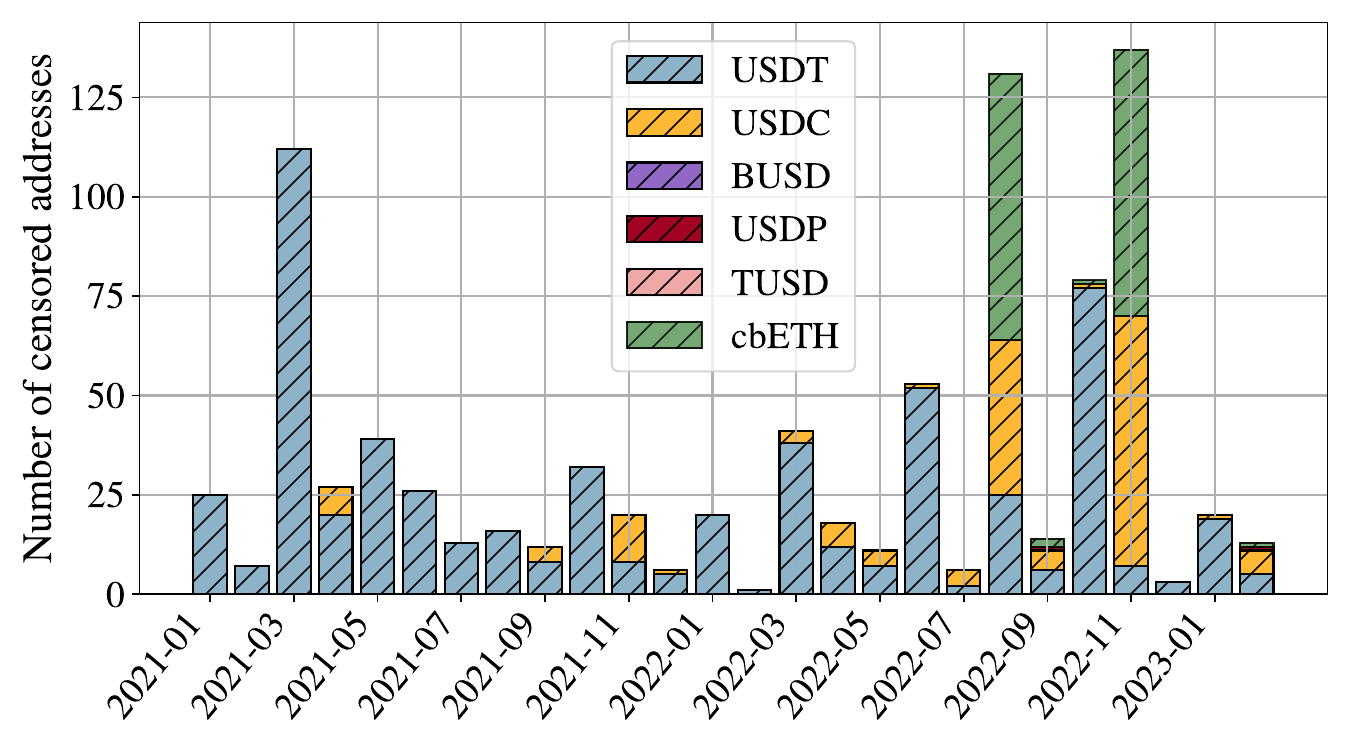}
\vspace{-2em}
\caption{Number of censored application layer accounts per month.
}
\label{fig:appcensoring}
\end{figure}

To quantify censorship at the application level (censorship by smart contract), we focus on a set of smart contracts that include functions to lock or freeze assets (cf.\ Figure~\ref{fig:appcensoring}). These contracts were deployed to the Ethereum blockchain but are controlled by the entity that deployed them, introducing trust requirements. Figure~\ref{fig:appcensoring} shows per month the number of newly censored addresses by these contracts. Between January $1^{st}$ and November $15^{th}$, the USDT contract blocked $556$ accounts. This exceeds the $87$ blocked accounts at USDC. For the stablecoins BUSD and TUSD we find that both have not blocked any address. For USDP, one account has been blocked, labelled as the ``\href{https://etherscan.io/address/0xe74b28c2eAe8679e3cCc3a94d5d0dE83CCB84705}{Wintermute Exploiter}'' on Etherscan, a \DeFi protocol that was exploited in September 2022 for $160$ million USD~\cite{faife2022hacker}.
For cbETH, we find that a total of $137$ accounts have been blocked since its deployment in February 2022.
Among the accounts blocked from interacting with the cbETH contract, we identify \TC's contracts as well as other \OFAC-sanctioned entities. In August 2022, when the sanctions against \TC were imposed, the number of blocked addresses reached $131$, corresponding to an increase of $84.99\%$ over the average number of blocked addresses per month from July 2021 to August 2022.

Censorship at the smart contract level can be implemented with \href{https://go.chainalysis.com/chainalysis-oracle-docs.html}{third-party smart contracts}. However, it is questionable how effective such solutions are, as the sanctioned entities could atomically forward assets to and use other accounts. 

\subsection{Censorship on Bitcoin}
\label{sec:BitcoinBlender}
There are several privacy-enhancing technologies on Bitcoin, such as centralized Bitcoin mixers and CoinJoin wallets~\cite{Wu_2021, Tironsakkul2022}. All were developed with the goal to prevent observers from tracing money flows through the ecosystem, enabling users to increase their on-chain privacy. In contrast to Ethereum, where shared addresses are used to obfuscate money flows, the UTXO-based Bitcoin blockchain relies on shared transactions among users. In the following, we exclusively focus on the centralized Bitcoin mixer Blender.io, since CoinJoin Wallets such as Wasabi Wallet or Samurai Wallet, were not targeted by \OFAC sanctions. Blender.io was sanctioned by \OFAC in May 2022, as discussed above.

\begin{figure}[!t]
\centering
\includegraphics[width=\columnwidth]{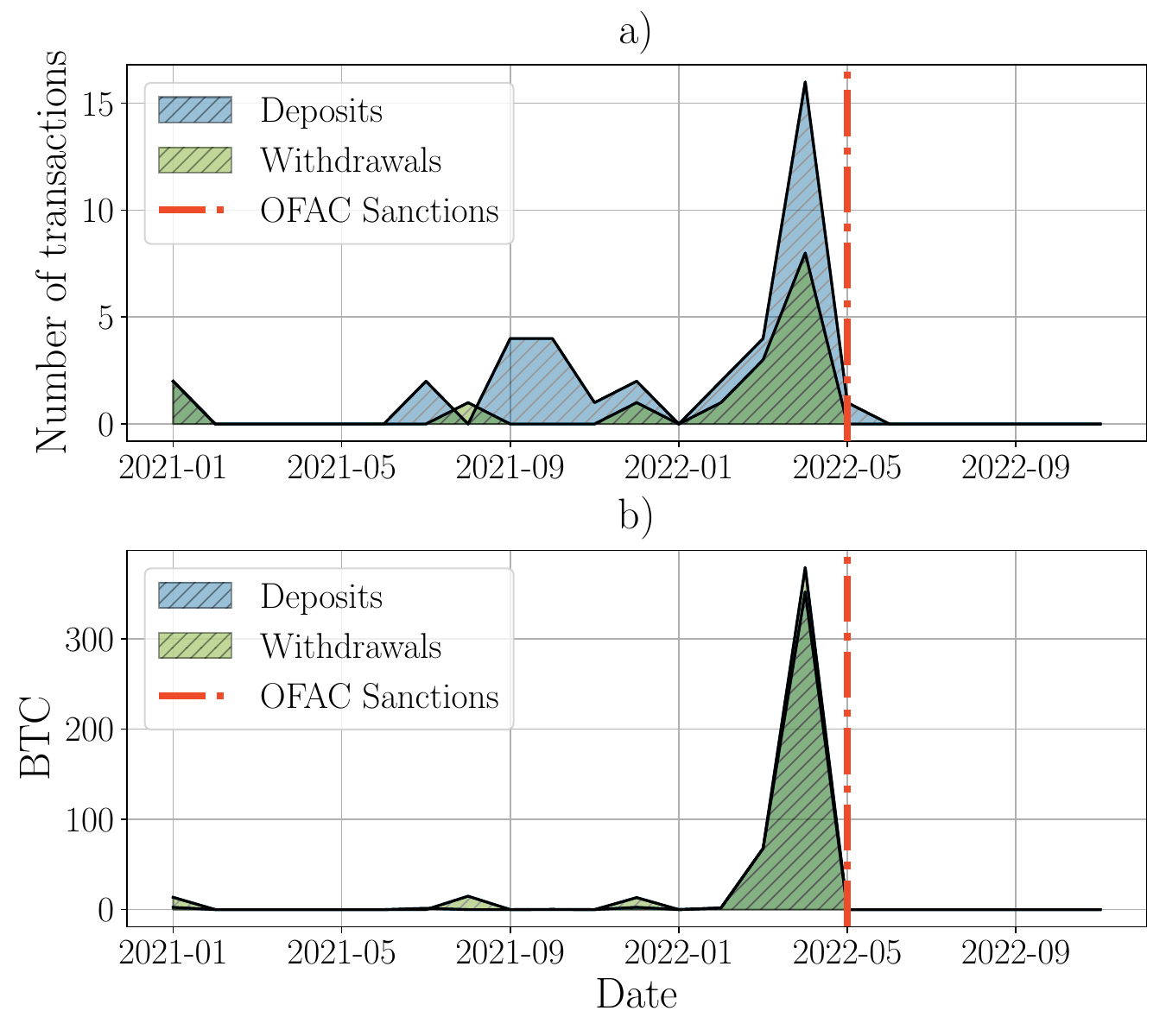}
\vspace{-2em}
\caption{Monthly interactions with the Bitcoin mixer Blender.io.
}
\label{fig:blender}
\end{figure}

In Figure \ref{fig:blender}, we visualize the interactions with the sanctioned Blender.io over time. Figure \ref{fig:blender} a) shows the number of transaction with Blender.io from January 2021 to November 15th 2022. In Figure \ref{fig:blender} b), we display the amount of BTC deposited and withdrawn for the same period. We observe that after \OFAC's sanctions against Blender.io were imposed, there were no further interactions with the application. Shortly before the sanctioning, there was an observable spike in both deposits to and withdrawals from Blender.io. We find that $351$ BTC and $379.06$ BTC were deposited and withdrawn, respectively, from addresses belonging to Blender.io in the month before the sanctioning. Assuming an exchange rate of $35000$ USD per BTC, around \$$10.5$ million were deposited and withdrawn in a single month, just before the sanctions took effect.

Figure~\ref{fig:blender} suggests that the sanctions fully prevented Blender.io from continuing to provide its centralized services. We do not propose that this occurred due to censorship as defined in Definitions~\ref{def:censorship}-\ref{def:contractcensorship}, as the likely cause is the removal of the Blender.io website.




\section{Security Implications of Censorship}
\label{sec:securityImplications}
In the following, we explore to what degree censorship affects blockchain security. In line with intuition, we find that censorship is slowing down transaction confirmation latency, which was shown to adversely affect double-spending resilience~\cite{karame2015misbehavior}. Finally, we share an impossibility result on censorship-resilience under \PoS.

\subsection{Transaction Confirmation Latency}\label{sec:confirmationLatency}
\begin{figure*}[htb]
\centering
\includegraphics[width=\textwidth]{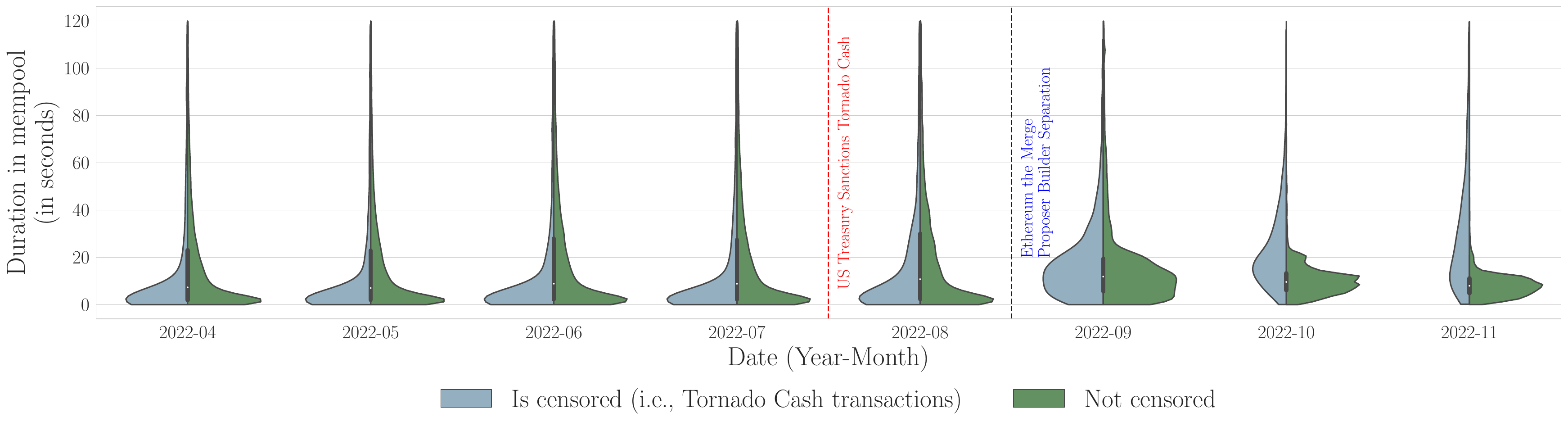}
\vspace{-1em}
\caption{Duration for transaction inclusions before and after the merge ($15th$~of September 2022), PBS ($15th$~of September 2022) and the sanctions ($8th$~of August 2022). We find that these events have a significant impact on transaction inclusion delay. For example, the average inclusion delay for \TC transactions increased from~$15.8 \pm 22.8$ seconds in August~$2022$ to~$29.3 \pm 23.9$ seconds in November~$2022$.}
\label{fig:tx-inclusion-duration}
\end{figure*}

In the following, we investigate the time it takes for transactions to be included in Ethereum, before and after the merge, while differentiating between censored and non-censored transactions (cf. Figure~\ref{fig:tx-inclusion-duration}).

\begin{description}[style=unboxed,leftmargin=0cm,labelindent=\parindent, itemsep=0.5em]
\item[Transaction Latency and Security.] Transaction confirmation latency matters from the security perspective due to the following reasons. First, research has shown that double-spending of zero-confirmation transactions is facilitated if a transaction remains longer in the mempool~\cite{karame2015misbehavior}. Second, under increased transaction confirmation latencies, sandwich attackers are more likely to successfully attack trading transactions~\cite{zhou2021high}. Third, price changes in automated market maker exchanges trigger transaction failures, if the trading transactions confirm ``slowly''~\cite{zhou2021a2mm}. Finally, systematically increased transaction latencies bear the risk of congesting the mempool, increasing the likelihood of transaction re-transmission and \PTwoP network layer congestion. Such congestion slows down the propagation of blocks and transactions, deteriorating blockchain security~\cite{gervais2015tampering,liu2022empirical}.

\item[Transaction Issuance Time.] We modify geth~\cite{zhou2021high,zhou2021a2mm}, currently the most popular client for Ethereum~\cite{ethernodes2022popularity}, to store all transactions received on the \PTwoP network layer between April 2022 and November 2022. Intuitively, the number of transactions a node can observe increases with the number of peer connections, the network bandwidth, and the computation power of the machine. Our geth node operates on an Ubuntu $20.04.2$ LTS machine with AMD Ryzen Threadripper 3990X (64-core, 2.9GHz), 256 GB of RAM and NVMe SSDs. We limit the geth client to at most~$1,000$ connections with other Ethereum peers instead of the default $50$ peers. The node is located in Europe. In total, we capture~$316.5$ million transactions in the aforementioned timeframe.

\item[Transaction Confirmation Time.] We rely on the timestamp data recorded in the block header as an estimate of the transaction confirmation time. It should be noted that this is a rough estimate of the confirmation time, because miners and validators (before and after the Ethereum merge, respectively) may decide not to report the precise timestamp that the blocks are generated at. For example, related works have identified evidence of miner misbehavior in block header timestamps for financial gain~\cite{yaish2022uncle}.

\item[Results.] After gathering the timestamps when a transaction emerges on the \PTwoP network and the time that the transaction is included on-chain, we can identify the relative time it takes for a transaction's inclusion. Further, we distinguish between transactions that are and are not subject to censorship (i.e., \TC and non-\TC transactions). To ensure a fair comparison, we only consider uncensored transactions that are mined in the same blocks as \TC transactions at a similar gas price (i.e., $\pm 10\%$). Two insights emerge:
\begin{enumerate}
    \item The time distribution of censored vs.\ non-censored transactions indicates that censored transactions remain in the mempool for a longer time and hence confirm slower on-chain. For example, as of November 2022, the average inclusion delay for non-censored transactions is~$8.7 \pm 8.3$ seconds, while the average inclusion delay for \TC transactions is~$29.3 \pm 23.9$ seconds.
    \item The inclusion latency of transactions has grown since Ethereum's transition to \PoS, and the adoption of PBS. For example, the average inclusion delay for \TC transactions increased from~$15.8 \pm 22.8$ seconds in August~$2022$ to~$29.3 \pm 23.9$ seconds in November~$2022$.
\end{enumerate}

\end{description}


\subsection{Supplementary measurement from an additional Ethereum node}
\label{sec:us_node} 
Blockchain mempool data appears to be heavily location-dependent. 
Due to network latency, a blockchain node may capture transactions late if they are submitted from a place physically far from the current node's location. 
For example, nodes in Europe might observe \TC transactions faster than others in America. 
In the worst case, a node may ``miss'' a transaction if  it is included swiftly in the blockchain. 
Relying on a single node could weaken the conclusions presented in \S~\ref{sec:confirmationLatency}. Therefore, we additionally collect data from a supplementary node in the US,  
and compare the results from our main node (in Europe) that we used in \S~\ref{sec:confirmationLatency}. 
We believe that having the mempool data from two different sources across continents corroborates our findings.

We evaluate the effect of a node's geographic location by comparing the transactions observed over a given timeframe. 
We collected transactions seen by our nodes for 15 days, from February 20th to March 7th, 2023. 
The supplementary US node uses \texttt{geth}'s default settings.
Our results are basically unchanged when raising the number of peer connections at the US node from 50 to 1,000.
Table~\ref{tab:two-nodes} summarizes the number of blockchain transactions each node observed.

\begin{table}[!t]
\centering
\caption{The number of blockchain transactions observed by each node.}
\label{tab:two-nodes}
\scalebox{0.90}{
\begin{tabular}{ccc}
\toprule
& \textbf{US observed}         & \textbf{US not observed} \\ \midrule
\textbf{Europe observed}            & 14,024,697 (96.1\%) & 58,742 (0.4\%)  \\
\textbf{Europe not observed}        & 86 (0.0\%)          & 505,465 (3.5\%) \\ \bottomrule
\end{tabular}
}
\end{table}

In total, 14,588,990 transactions were included in Ethereum blocks over our data collection period, and both nodes observed more than 95\% of these transactions, before their inclusion in a block. 
Our (main) European node observed some transactions (about 4,000 per day) that the US node did not. 
On the contrary, only 0.0005\% of 
transactions were only observed by the supplementary US node---and not by the European node.
Our results imply that our main node is more powerful than the supplementary node. 
Also, both nodes failed to observe 3.5\% of all transactions. 
These are likely to be private transactions directly submitted to validators without going through the public mempool~\cite{qin2022quantifying}. 
The number of private transactions seems to be about the same as the study by Lyu~\emph{et al.}~\cite{lyu2022empirical}, but this ratio is much higher for \TC transactions (16.8\%).

We also calculate the inclusion time for both nodes (i.e., the difference between the time a node observes a transaction, and the timestamp of the block the transaction was eventually included in). 
More than 99\% of the time, the observed inclusion time for the Europe node is lower than that for the US node.
The difference is lower than $1$ second for more than 99\% of observed transactions, and under $0.1$ seconds for 30\% of them. 
The difference is statistically significant based on the paired $t$-test (p-value: $\ll$0.01). 
The European node and the US node have slightly different ways of recording transactions, which may have affected the time inclusion. The European node records the transaction before it gets into the mempool, whereas the US one observes just after it hits the mempool. That may have caused the Europe one to hear the transaction first.

We also compare transaction observation time \textit{between} the two nodes and check if there is a statistically significant difference between \TC and non-\TC transactions. In other words, we test if one node observes \TC transactions relatively faster than the other.
The result suggests that the difference is statistically insignificant across \TC and non-\TC transactions, based on the t-test (p-value: 0.116). 
Furthermore, for both nodes, \TC transactions take more time to be included than non-\TC transactions with a significant margin (p-value: $\ll$0.01), thereby corroborating the conclusions of \S~\ref{sec:confirmationLatency}. 
This time, we did not control for confounding factors such as gas fees.
For all the statistical tests discussed here, we set the threshold to 5\% and excluded outliers (i.e., transactions that take more than 120 seconds to be included in a block).  
In short, the observations given by our main node are reliable enough to conduct all the analyses of \S~\ref{sec:confirmationLatency}. 

\subsection{Impossibility of Censorship-Resilience}\label{sec:impossibility}
In this section, we argue that previous results on liveness in \PoS~\cite{tas2022bitcoin, gao2022dumbo, sheng2021bft} constitute a lower bound for censorship-resilience on the consensus layer. 
Concretely, we prove that no \PoS protocol can achieve \textit{censorship-resilience}, if the number of censoring validators makes up more than 50\% of the validator committee. 
In the following, we first introduce our model in reasoning about security in \PoS blockchains. We outline recent results on \textit{liveness} in \PoS blockchains and further introduce the relationship of censorship-resilience to liveness. After introducing an intuition, we state our impossibility result in Theorem~\ref{theorem:impossibility} and prove it through a sequence of worlds and an indistinguishability argument.


\begin{description}[style=unboxed,leftmargin=0cm,labelindent=\parindent, itemsep=0.5em]

    \item[Security Model. ]
    Recall that $\blockShort \in \ledgerShort$, if a block $\blockShort$ is included in the distributed ledger $\ledgerShort$, and that $\transactionShort$ is included  in $\ledgerShort$ if $\transactionShort \in \blockShort \wedge \blockShort \in \ledgerShort$ (cf. Section~\ref{section:notation}). Two views of ledgers $\ledgerShort_1, \ledgerShort_2$ are conflicting if they differ in their included transactions.
    We further assume that $\numberValidators$ is the total number of validators. 
    In our model, transactions $\transactionShort$ are input to validators by the environment $\environmentShort$.
    Before the execution of the protocol starts, an adversary $\adversaryShort$ corrupts a subset of validators $\numberCorruptedValidators < \numberValidators$ and renders them \textit{adversarial} such that they can arbitrarily deviate from the specified protocol. Remaining validators are \textit{honest} and follow the protocol as specified.
    We assume that network communication is synchronous, hence messages are instantly delivered once they are sent by a network participant.

    \item[Safety \& Liveness. ]
    The safety and liveness of Proof-of-Stake blockchains was studied under varying synchronicity assumptions~\cite{gavzi2019proof, tas2022bitcoin, gao2022dumbo}, and assuming dynamic validator committees~\cite{neu2022availability}.  However, definitions of liveness subtly differ in their phrasing.
    Whereas some works focus on the eventual inclusion of a transaction in a block after a finality delay $\Delta$~\cite{miller2016honey, neu2022availability, tas2022bitcoin}, others focus on the correct report of a value upon a query sent by an honest client~\cite{gavzi2019proof}.
    To formally define liveness, we follow the holistic definition of Garay~\emph{et al.}~\cite{garay2015bitcoin}, which states that all transactions originating from an honest client will eventually end up in an honest validators' view of a ledger, hence an adversary cannot perform a DoS attack against honest clients.
    We formally define properties of liveness in a \PoS protocol as follows.
    \begin{definition}
        A validator ensures \textit{liveness} of a \PoS protocol, if it satisfies the following properties:
        \begin{enumerate}
            \item \textbf{Propagation. } Upon reception of a transaction $\transactionShort$ by an honest client $\clientShort$, the validator forwards $\transactionShort$ to other peers in the network.
            \item \textbf{Inclusion. }
            A transaction $\transactionShort$ sent by an honest client $\clientShort$ is eventually included in the local view of an honest validator's distributed ledger $\ledgerShort$.
            \item \textbf{Availability. } Upon query, an honest validator will report whether a transaction is included in the ledger.
        \end{enumerate}
    \end{definition}
    Importantly, recent results highlight a trade-off between accountability and availability~\cite{neu2022availability} and show the impossibility of liveness beyond $\fractionAdversarialValidators > \frac{n}{2}$ adversarial validators~\cite{tas2022bitcoin}.
    
    \item[Modeling Censorship. ]
    In a real-world environment, a validator censoring transactions may not be considered adversarial.
    For example, censorship may be considered beneficial from a legal perspective, as malicious actors are prevented from participating in the system.
    However, as first identified by Miller~\emph{et al.}~\cite{miller2016honey}, censorship-resilience is a property of liveness. We further argue that the act of censorship is equivalent to a subset of adversarial actions as defined in a \PoS protocol.
    To formally define this finding, we say that $\adversaryShortCR$ is a probabilistic polynomial time algorithm, where a subset $\numberCensoringValidators < \numberValidators$ of validators, which are corrupted by $\adversaryShortCR$, arbitrarily deviate from the \PoS protocol. For example, a censoring validator may, upon reception of a transaction $\transactionShort$ by the environment, $\environmentShort$ by refusing to \emph{(i)} include $\transactionShort$ in block $\blockShort$, \emph{(ii)} propagate $\transactionShort$ to other peers in the network \emph{(iii)} build upon $\ledgerShort$, where $\transactionShort$ is included in $\ledgerShort$, and by \emph{(iv)} refusing to attest to $\blockShort$, where $\transactionShort \in \blockShort$. 
    Given the previous reasoning, we define \textit{$\Delta$-censorship-resilience} as follows.

    \begin{definition} ($\Delta$-Censorship-resilience)
        Suppose an honest client inputs $\transactionShort$ to $(\numberValidators - \numberCensoringValidators)$ honest validators. Then,  $\transactionShort$ is committed to $\ledgerShort$ within $\Delta$, except with negligible probability. 
    \end{definition}


    \item[Intuition. ]
    Let us consider a \PoS protocol where $\numberCensoringValidators > \frac{n}{2}$ of validator nodes are directly censoring transactions.
    We show that censoring impacts the liveness of a blockchain. Intuitively, censorship of a transaction prevents it from being included in the blockchain, as a censoring validator drops transactions that, e.g., involve sanctioned addresses.
    As such, censoring validators create a conflicting chain, which can be considered \textit{adversarial} in the context of traditional Byzantine Fault Tolerant consensus protocols.
    To prove Theorem 1, we show that the threat to liveness posed by a validator corrupted by $\adversaryShort$ is indistinguishable from the threat to liveness posed by a validator corrupted by $\adversaryShortCR$, hence that are censoring. We defer the proof of liveness to the argument presented by Tas~\emph{et al.}~\cite{tas2022bitcoin} and present Theorem~\ref{theorem:impossibility}.







    %
    
\end{description}

\begin{theorem}
\label{theorem:impossibility}
Consider a \PoS protocol $\protocolShort$ with $\numberValidators$ validators in a synchronous network, where at least $\numberCensoringValidators > \frac{n}{2}$ are corrupted by $\adversaryShortCR$. Then, $\protocolShort$ cannot provide \textit{censorship-resilience}.
\end{theorem}

\begin{description}[style=unboxed,leftmargin=0cm,labelindent=\parindent, itemsep=0.5em]
    \item[Proof. ]
    Suppose the number of validators is $\numberValidators \in \environmentShort$, where we assume that $\numberValidators$ is even in each epoch. 
    Further, consider there exists a protocol $\protocolShort$ that supports \textit{liveness} with $\numberCorruptedValidators < \frac{n}{2}$ corrupted validators that is further \textit{$\Delta$-censorship-resilient} with $\numberCensoringValidators > \frac{n}{2} - \numberCorruptedValidators$ censoring validators.
    Then, there exists a decision function $\decisionFunctionShort$, which outputs a non-empty set of censoring validators.
    We prove Theorem~\ref{theorem:impossibility} through a sequence of worlds and an indistinguishability argument. \\
    \textit{(World 1.) }
    Let $\groupOne$, $\groupTwo$ and $\groupThree$ partition $\numberValidators$ into three disjoint groups, where $\abs{\groupOne} < \frac{n}{2} , \abs{\groupTwo} > \frac{n}{2} - \abs{\groupOne}$ and $\groupThree = \numberValidators - \abs{\groupOne} - \abs{\groupTwo}$.
    Nodes in $\groupOne$ are corrupted by $\adversaryShort$, nodes in $\groupTwo$ are corrupted by $\adversaryShortCR$ and nodes in $\groupThree$ are honest.
    Corrupted nodes are adversarial and do not communicate with honest nodes in $\groupThree$.
    Hence, upon input of randomly distributed transactions by $\environmentShort$, validators in $\groupOne \cup \groupTwo$ output a diverging view of $\ledgerShort$ as opposed to $\groupThree$.
    So the decision function outputs a non-empty set of adversarial validators $\groupOne \cup \groupTwo$.  
    \\
    \textit{(World 2.) }
    Let $\groupOne$, $\groupTwo$ and $\groupThree$ partition $\numberValidators$ into three disjoint groups, where $\abs{\groupOne} < \frac{n}{2} , \abs{\groupTwo} > \frac{n}{2} - \abs{\groupOne}$ and $\groupThree = \numberValidators - \abs{\groupOne} - \abs{\groupTwo}$.
    Nodes in $\groupOne$ are corrupted by $\adversaryShortCR$, nodes in $\groupTwo$ are corrupted by $\adversaryShort$ and nodes in $\groupThree$ are honest.
    Corrupted nodes are adversarial and do not communicate with honest nodes in $\groupThree$.
    Hence, upon input of randomly distributed transactions by $\environmentShort$, validators in $\groupOne \cup \groupTwo$ output a diverging view of $\ledgerShort$ as opposed to $\groupThree$.
    So the decision function outputs a non-empty set of adversarial validators $\groupOne \cup \groupTwo$.

    




    
\end{description}

However, worlds $1$ and $2$ are indistinguishable for the decision function $\decisionFunctionShort$. Thus, $\decisionFunctionShort$ cannot output a non-empty set of censoring validators. 

    \hfill $\square$

We note that this lower bound for censorship-resilience also applies to Nakamoto consensus~\cite{nakamoto2008bitcoin,garay2015bitcoin}.

\section{Discussion}\label{sec:discussion}
Our quantitative results suggest that it is not so trivial to treat individual actors separately from each other, as the behaviour of each one may affect the practices of others. 
Block proposers who use MEV-Boost depend on block builders and relayers to serve them the payloads for their blocks. Following, block proposers, who by default use a profit-maximizing strategy to choose payloads, automatically accept the block assigned to them without verifying if when doing so they contribute to censorship. By building blocks locally (as was the standard, before \PBS), or by exclusively connecting to non-censoring relays, block proposers can ensure to not partake in censorship. Furthermore, proposers can use the \textit{min-bid} flag offered by the MEV-Boost software that enables them to automatically fall back to uncensored block building (``vanilla'' building) if the payments offered by blocks constructed by builders are not above a certain threshold~\cite{hu2022cost}.

For external block builders, who want to maximize the number of blocks they create that are added to the blockchain, censorship by relayers may push them towards producing censored blocks. For example, if the majority of block proposers are exclusively connected to a single relay, builders are forced to comply with the censoring practices of the respective relay to get their block payloads delivered. Similar to the well-known \textit{feather forking} attack, the financial profit that can be gained by creating blocks that conform with the censorship practices of relayers can pressure external block builders and MEV searchers to comply, as well. 

With respect to execution delays, we argue that the consequences depend heavily on the individual application. For \TC, a delay in the inclusion and execution of a transaction may not have significant impact on the user experience or the contract itself, because there are no deadlines within the contract that might cause transactions to revert if they are not incorporated into a block in a timely manner. For more time-dependent contracts such as decentralized exchanges, execution delays may have a significant impact on user experience. For security reasons, many DeFi contracts include functionalities that protect users from executing transactions that ultimately entail considerable disadvantages in terms of execution price because too much time has passed between the desired and the actual execution. So, despite the fact that the current censorship practices might not have a significantly large impact on the end user, it strongly depends on the individual application.

Furthermore, we show that \PoS consensus protocols cannot achieve censorship-resilience in scenarios in which more than 50\% of the nodes adopt censorship practices. Censoring nodes, if they are the majority, can exclusively create blocks that do not contain interactions with sanctioned entities, while at the same time abstaining from attesting to blocks created by peers that do not adhere to the same rules.
Thus, the only blocks that are appended to the blockchain are those adhering to their rules.

\section{Conclusion}\label{sec:conclusion}

In this paper, we investigate the impact of censorship in blockchains. We present a systematization of blockchain censorship through formal definitions and through a quantification of the effect of \OFAC sanctions on censorship in Ethereum and Bitcoin. We find that after transitioning to \PoS, $46\%$ of Ethereum blocks were made by censoring actors that intend to comply with \OFAC sanctions. Additionally, we reason about their impact on blockchain security. We find that censorship prolongs the time until a transaction is confirmed, which degrades blockchain security. Finally, we prove that if $> 50\%$ of validator nodes enforce censorship, a blockchain cannot be censorship resilient. 

Our results show that censorship on blockchains is not a mere hypothetical threat: it already degrades the security of existing blockchains, and the quality of service for users. Our work sheds light on a dilemma anticipated for a decade: if regulators intervene, will the promise of a permissionless, secure append-only ledger withstand?
We hope that this work draws attention to the significance of censorship in permissionless blockchains and engenders future work on addressing the mentioned security issues. 

\bibliographystyle{ACM-Reference-Format}
\bibliography{references}

\end{document}

%% file: macros.tex
\DeclareAcronym{PBS}{
  short = PBS,
  long  = Proposer-Builder Separation,
}
\newcommand{\PBS}{\ac{PBS}\xspace}

\newcommand{\blockchain}{blockchain}

\newcommand{\consensusLayer}{consensus layer\xspace}
\newcommand{\ConsensusLayer}{Consensus Layer\xspace}

\newcommand{\PTwoP}{P2P\xspace}

\newcommand{\blockHeightShort}{h}
\newcommand{\blockIdentifierShort}{i}
\newcommand{\blockShort}{B_{\blockIdentifierShort}}

\newcommand{\numberValidators}{n}
\newcommand{\numberCorruptedValidators}{f}
\newcommand{\numberCensoringValidators}{f_c}

\newcommand{\adversaryShort}{\mathcal{A}}
\newcommand{\adversaryShortCR}{\adversaryShort_{C}}
\newcommand{\transactionShort}{tx}
\newcommand{\transactionsShort}{\mathbf{tx}}
\newcommand{\clientShort}{\mathcal{C}}
\newcommand{\environmentShort}{\mathcal{Z}}
\newcommand{\ledgerShort}{\mathcal{L}}
\newcommand{\protocolShort}{\Pi}
\newcommand{\fractionAdversarialValidators}{f}
\newcommand{\decisionFunctionShort}{\mathcal{D}}
\newcommand{\groupOne}{P}
\newcommand{\groupTwo}{Q}
\newcommand{\groupThree}{R}

\DeclareAcronym{SMR}{
  short = SMR,
  long  = State Machine Replication,
}
\newcommand{\stateMachineReplication}{\ac{SMR}\xspace}
\newcommand{\stateMachineReplicationShort}{\Pi}

\newcommand{\finalityTime}{T_{\Delta}}
\newcommand{\securityParameter}{\sigma}

\newcommand{\addressShort}{a_i}
\newcommand{\stateShort}{st}

\newcommand{\direct}{direct\xspace}

\newcommand{\directly}{directly\xspace}
\newcommand{\indirectly}{indirectly\xspace}